\documentclass[aps,prl,twocolumn,superscriptaddress,nofootinbib]{revtex4}

\usepackage{amsmath,amssymb,graphicx,hyperref}
\usepackage{xcolor}
\usepackage[utf8]{inputenc}
\usepackage{booktabs}
\usepackage{placeins}
\usepackage{float}
\usepackage[normalem]{ulem}

\newcommand{\OO}{\ensuremath{\mathcal{O}}}
\newcommand{\Op}[1]{\OO_{\sss #1}}
\newcommand{\sss}{\scriptscriptstyle}
\newcommand{\pdp}{\ensuremath{\varphi^\dagger\varphi}}
\newcommand{\ccc}[3]{c_{#2}^{#1 (#3)}}

\def\lra#1{\overset{\text{\scriptsize$\leftrightarrow$}}{#1}}

\begin{document}

\title{Data-driven discovery strategy for standard model effective field theory searches}

\author{Martin Hirsch}
\author{Luca Mantani}
\author{Veronica Sanz}
\affiliation{Instituto de F\'isica Corpuscular (IFIC), Universidad de Valencia-CSIC, E-46980 Valencia, Spain}

\begin{abstract}
We present a novel strategy to uncover indirect signs of new physics in collider data using the Standard Model Effective Field Theory (SMEFT) framework, offering notably improved sensitivity compared to traditional global analyses. Our approach leverages genetic algorithms to efficiently navigate the high-dimensional space of operator subsets, identifying deformations that improve agreement with data without relying on prior UV assumptions. This enables the systematic detection of SMEFT scenarios that outperform the Standard Model in explaining observed deviations. We validate the approach on current LHC and LEP measurements, perform closure tests with injected UV signals, and assess performance under high-luminosity projections. The algorithm successfully recovers relevant operator subsets and highlights directions in parameter space where deviations are most likely to emerge. Our results demonstrate the potential of SMEFT-based discovery searches driven by model selection, providing a scalable framework for future data analyses.
\end{abstract}

\maketitle

\noindent {\bf Introduction.}
Despite its success, the Standard Model (SM) of particle physics fails to explain key phenomena such as dark matter, neutrino masses, and the baryon asymmetry of the universe. These shortcomings motivate the search for new physics (NP), especially at the LHC and other high-energy experiments, where deviations from SM predictions could signal physics beyond the current paradigm.

In scenarios where NP lies at energy scales higher than those directly accessible by current experiments, the Standard Model Effective Field Theory (SMEFT) offers a systematic and model-independent framework to parameterize possible deviations from SM expectations. This is achieved by extending the SM Lagrangian with higher-dimensional operators built from SM fields, each weighted by a corresponding Wilson coefficient and organized by their canonical mass dimension~\cite{Brivio:2017vri,Isidori:2023pyp,Aebischer:2025qhh}. Phenomenological analyses typically focus on dimension-six operators~\cite{Grzadkowski:2010es}, which represent the leading-order corrections under the assumption of lepton number conservation.

In practice, probing the SMEFT parameter space is highly challenging: depending on the underlying flavour assumptions, the number of independent dimension-six operators can reach into the hundreds. Moreover, fully constraining these operators requires combining a large and diverse set of experimental observables from different processes and energy scales. Traditional studies have approached this problem through global SMEFT fits~\cite{de_Blas_2020,Brivio_2022,Ellis_2021,Cirigliano:2016nyn,Cirigliano:2023nol,Garosi:2023yxg,Bissmann:2019gfc,Bissmann:2020mfi,Allwicher:2023shc,Bruggisser:2021duo,Bruggisser:2022rhb,Bellafronte:2023amz,Grunwald:2023nli,Hiller:2024vtr,Hiller:2025hpf,Hoeve:2025yup,Mantani:2025bqu,terHoeve:2025gey,Celada:2024mcf,Maura:2025rcv,deBlas:2025xhe}, often restricting the analysis to a limited set of operators or focusing on specific UV-inspired scenarios.

In this work, we go beyond the traditional approach by proposing an alternative strategy to boost discovery potential. Our goal is to agnostically identify statistically significant deviations from SM predictions that may signal the presence of NP, as captured by particular operator combinations, without assuming correlations among Wilson coefficients or imposing UV-inspired priors. To tackle this challenge, we employ genetic algorithms~\cite{Holland:1975,goldberg89}, stochastic optimization methods inspired by biological evolution, that efficiently explore subsets of the high-dimensional SMEFT parameter space. To discriminate genuine NP signals from statistical fluctuations, we rely on established model selection criteria, such as the Bayesian Information Criterion (BIC), a large sample approximation of the Bayes factor~\cite{Schwarz:1978tpv}, and the Akaike Information Criterion (AIC)~\cite{Akaike1974AIC}. While similar agnostic strategies have been proposed for model-independent NP searches using hypothesis testing and anomaly detection with machine learning~\cite{DAgnolo:2018cun,DAgnolo:2019vbw,dAgnolo:2021aun,Letizia:2022xbe,Grosso:2023scl,Grosso:2024nho,Grosso:2024wjt,Khosa:2020qrz,CMS:2020zjg,ATLAS:2018zdn,Belis:2023mqs}, our approach is distinguished by its tailored formulation within the SMEFT paradigm.  A related methodology employing the AIC as a model selection criterion
to reduce the dimensionality of SMEFT fits has been proposed in Ref.~\cite{Cirigliano:2023nol}. Their method toggles small groups of operators on and off, yielding only a limited number of combinations, in contrast to our genetic algorithm strategy. 

Closure tests on synthetic data indicate that our method can identify NP signals when present and avoids false positives under the SM, providing a solid basis for application to real data. Moreover, our framework opens a promising avenue for assessing the physics discovery potential of future colliders such as the FCC~\cite{Benedikt:2928193}, the ILC/LCF~\cite{LinearColliderVision:2025hlt}, CEPC~\cite{CEPCPhysicsStudyGroup:2022uwl}, CLIC~\cite{Aicheler:2012bya}, and the Muon Collider~\cite{InternationalMuonCollider:2024jyv}.

\noindent {\bf Methodology.}
For our analysis, we employ the \texttt{smefit} framework~\cite{Giani_2023}, which provides a consistent implementation of dimension-six operator effects across a broad range of collider observables. We consider two scenarios: a present-day dataset (LHC), comprising 442 data points, and a high-luminosity projection (HL-LHC), which extends this to 698. Both datasets include electroweak precision observables (EWPOs) from LEP, along with Higgs, top-quark, and diboson measurements from LHC Run~2. The HL-LHC scenario~\cite{Celada:2024mcf} further incorporates anticipated improvements in experimental precision and extends the dataset with additional processes accessible at the upgraded collider, such as double Higgs production~\cite{Atlas:2025kye} and higher invariant mass bins in $t\bar{t}$ distributions~\cite{Durieux:2022cvf}.

We consider a set of 52 dimension-six SMEFT operators, including all CP-even terms relevant to the observables studied (see the End Matter for definitions). Renormalization group evolution (RGE) effects are systematically included~\cite{Jenkins:2013wua,Jenkins:2013zja,Alonso:2013hga}, as they are essential to consistently connect the high-energy matching scale, where SMEFT operators arise from UV completions~\cite{terHoeve:2023pvs}, to the lower-energy scales probed experimentally~\cite{terHoeve:2025gey,Bartocci:2024fmm}.   Additional operators generated through mixing, such as several four-lepton and two-lepton–two-quark interactions, do not contribute to any of the considered observables at tree level and are therefore omitted~\cite{terHoeve:2025gey}. In this analysis, Wilson coefficients are defined at the reference scale $\mu_0 = 1\,\text{TeV}$. Importantly, RGE induces operator mixing, allowing observables to receive contributions from operators not directly generated at the matching scale~\cite{Battaglia:2021nys,Aoude:2022aro,DiNoi:2023onw,Garosi:2023yxg,Bartocci:2024fmm,Allwicher:2023shc,Maltoni:2024dpn,Greljo:2023bdy,Duhr:2025zqw}. This is particularly relevant for EWPOs at the $Z$-pole, where current and anticipated precision, e.g., at FCC-ee~\cite{Allwicher:2024sso}, make such loop-induced effects a powerful probe of NP.

To quantify how well different SMEFT extensions describe the data, we define each ``model'' as a specific subset of the 52 dimension-six operators considered and evaluate its goodness of fit by minimizing the corresponding chi-squared statistic:
\[
\hat{\chi}^2 = \left( \mathbf{D} - \mathbf{T}^{\text{EFT}} \right)^{\!\!T}
\, \mathbf{\Sigma}^{-1} \left( \mathbf{D} - \mathbf{T}^{\text{EFT}} \right),
\]
where $\mathbf{D}$ and $\mathbf{T}^{\text{EFT}}$ denote the vectors of experimental measurements and SMEFT predictions, respectively, and $\mathbf{\Sigma}$ is the total covariance matrix. This matrix accounts for both experimental uncertainties, statistical and systematic, as well as theoretical uncertainties from missing higher-order corrections in the SM and parton distribution function errors.
We retain only the linear contributions of dimension-six operators, neglecting quadratic terms in the Wilson coefficients.
This approximation is justified as long as the Wilson coefficients remain sufficiently small for quadratic corrections to be subdominant, which holds for many operators under current experimental bounds, though some directions probe the nonlinear regime.
Under this assumption, the theory predictions take the form
\begin{equation}
    \mathbf{T}^{\text{EFT}} = \mathbf{T}^{\text{SM}} + \sum_{i=1}^{n_{\text{op}}} \frac{c_i(\mu_0)}{\Lambda^2} \, \mathbf{T}_i^{\text{int}} \,,
\end{equation}
where $\mathbf{T}^{\text{SM}}$ denotes the SM prediction and $\mathbf{T}_i^{\text{int}}$ encodes the interference between the SM and the $i$-th dimension-six operator.

To account for model complexity and implement Occam’s razor, we evaluate two model selection criteria:
\[
\mathrm{AIC} = \hat{\chi}^2 + 2 \, n_{\text{op}}, \qquad \mathrm{BIC} = \hat{\chi}^2 + n_{\text{op}} \, \ln N_{\text{dat}}\,,
\]
where $n_{\text{op}}$ is the number of active parameters (i.e., operators) and $N_{\text{dat}}$ is the number of data points. Both criteria penalize overly complex models to avoid overfitting, with the BIC introducing a stronger penalty that increases with the dataset size. In our setup, we find that BIC typically offers more robust discrimination against noise-driven solutions.

To efficiently explore the exponentially large model space, $d = 2^{n_{\text{op}}} \approx 10^{15}$, we use a genetic algorithm (GA)~\cite{gad2021pygadintuitivegeneticalgorithm} where each candidate model is encoded as a binary string indicating operator presence. The GA evolves a population of models via selection, crossover, and mutation, guided by the statistical estimators introduced above.  More details on its implementation are reported in the End Matter material.

Our implementation evolves a population of 2000 models over 200 generations. In each generation, 100 parents are selected via tournament selection (tournament size $K=4$), offspring are generated using uniform crossover, and each bit is subject to a mutation probability of $2.5\%$. We have verified the robustness of our conclusions under moderate variations of these hyperparameters.

To assess the sensitivity and reliability of our method, we define two levels of synthetic data, generated under either the SM or a specific BSM hypothesis. In Level~0 (L0), pseudodata exactly match the central theory predictions, without statistical fluctuations. In Level~1 (L1), pseudodata are sampled from a multivariate Gaussian centered on the theory prediction, using the experimental covariance matrix to model noise.   Note that when extending the dataset to the HL-LHC, statistical fluctuations are independent of those from the LHC. This setup allows us to benchmark both false positives (from SM generation) and discovery potential (from BSM injections), and to disentangle the algorithm’s intrinsic resolution from statistical limitations imposed by data fluctuations.

\noindent {\bf Standard Model closure test.}
\begin{figure}
    \centering
    \includegraphics[width=1.0\linewidth]{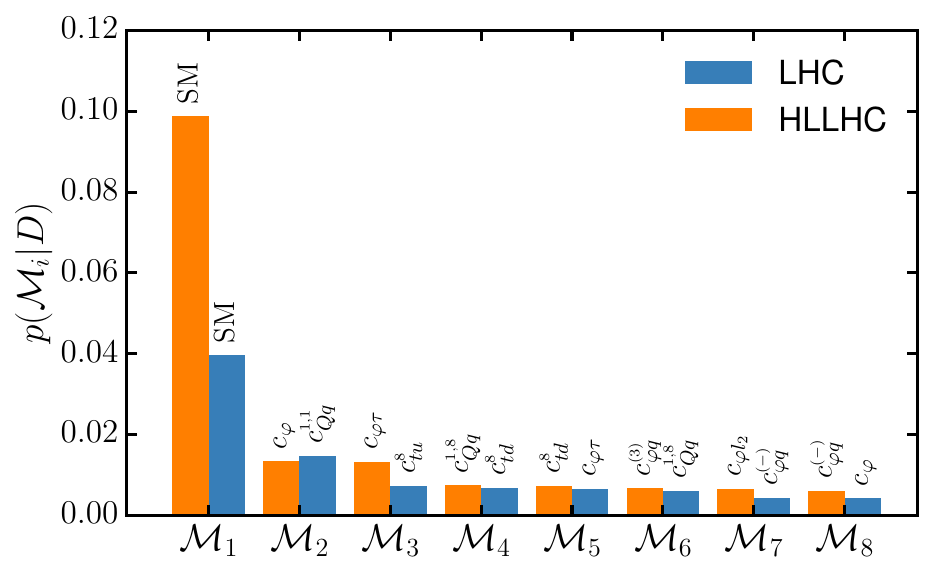}
    \caption{Bar plot showing the pseudo-posterior probabilities of the 8 top-ranked models based on their BIC scores, in a closure test with L1 SM-generated pseudodata. The SM emerges as the most probable model, with its dominance increasing at the HL-LHC as expected from improved precision.}
    \label{fig:sm-closure}
\end{figure}
\begin{figure*}[t]
    \centering
    \includegraphics[width=0.49\textwidth]{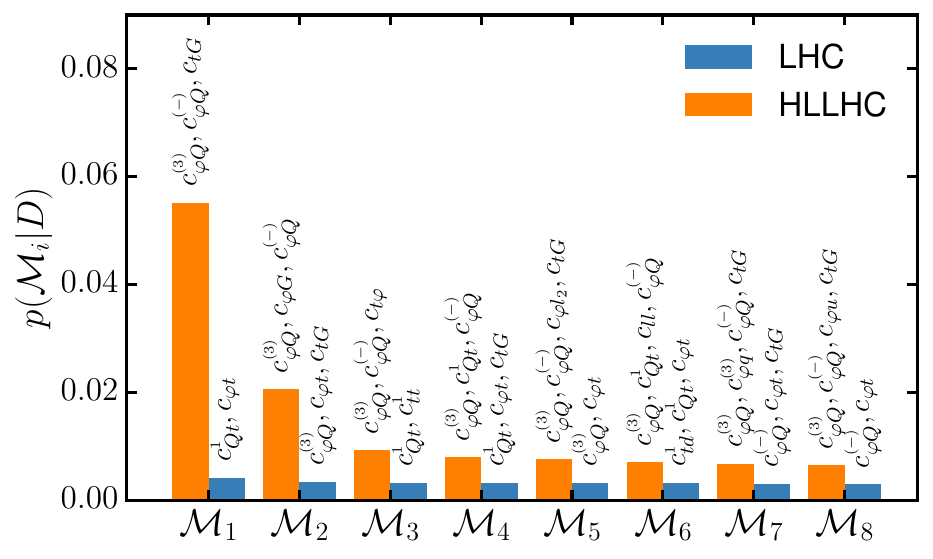}
    \hfill
    \includegraphics[width=0.49\textwidth]{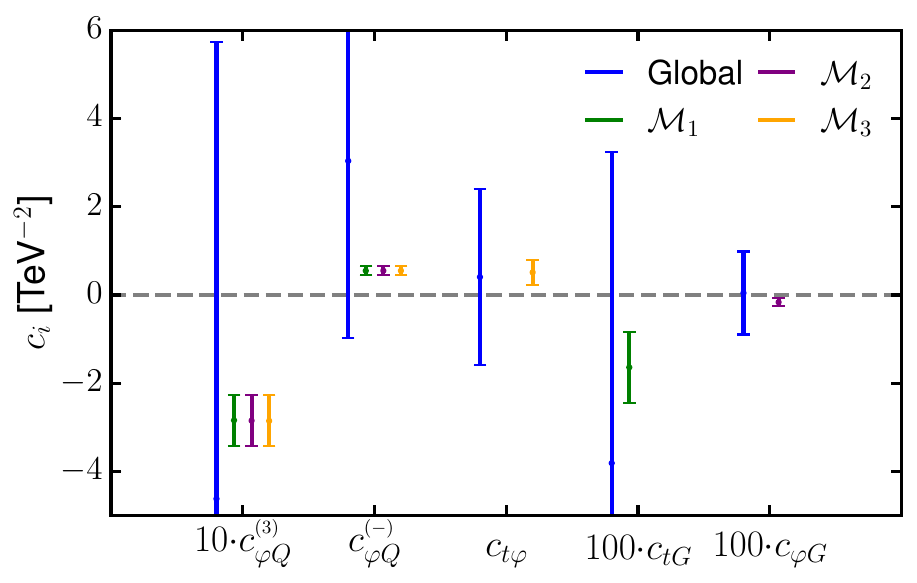}
    \caption{Left: Same as Fig.~\ref{fig:sm-closure}, but for BSM-injected pseudodata. At HL-LHC, two injected operators are reliably recovered, while the Yukawa-like one remains elusive due to degeneracies. At LHC precision, the SM is disfavoured, yet the characterization of NP remains inconclusive. Right: Comparison of the 95\% confidence intervals for selected operators between the top three scoring models and the global fit, using the HL-LHC dataset. The result highlights that the global fit fails to accurately characterize the BSM signal.}
    \label{fig:bsm-closure}
\end{figure*}
We begin by validating our method on pseudodata generated under the SM hypothesis, considering both the L0 and L1 scenarios. In the L0 case, where data exactly match the SM predictions, the genetic algorithm consistently identifies the SM as the optimal solution with $\chi^2 = 0$, and no improvement is gained by activating additional operators. This outcome is analytically expected: models with extra operators contain the SM as a limiting case and can reproduce the data equally well. Model selection criteria then serve only to penalize complexity, ranking models by parameter count. We confirm that the algorithm reproduces this hierarchy, with the SM always selected as best-fit and the score landscape correctly structured. This demonstrates that, in the absence of statistical fluctuations, our method behaves as expected and faithfully captures the complexity structure of the theory space.

In the L1 scenario, we generate 10 pseudodata replicas by fluctuating the SM predictions according to the experimental covariance matrix. In this setting, the genetic algorithm occasionally identifies operator subsets that achieve a moderate reduction in $\chi^2$ by fitting statistical noise. However, these improvements are typically not substantial enough to overcome the complexity penalties imposed by model selection criteria. When using the AIC, we observe in all data replicas a tendency to favor more complex EFT models with 5–10 active operators, yielding moderate preferences over the SM with $\Delta\mathrm{AIC} \sim 5$--10.

In contrast, the BIC shows more conservative behavior, closely reproducing the ideal model hierarchy seen in the L0 test. In 8 out of 10 L1 replicas, it selects the SM as the best model. In the remaining cases, the favored models contain only one or two operators, with $\Delta\mathrm{BIC} \lesssim 2$ with respect to the SM. Across the model population, BIC consistently ranks simpler combinations higher, demonstrating its robustness against false positives and its fidelity to model parsimony under fluctuations.

We repeat the L1 closure test using the extended dataset with HL-LHC projections, increasing the observables from 442 to 698. As expected, the SM becomes more favored: greater statistical power reduces the impact of fluctuations, allowing model selection criteria to penalize complexity more effectively. This aligns with Bayesian reasoning~\cite{Trotta:2008qt}, as more data concentrate the posterior around the true model.

To quantify the statistical preference among models, we define a pseudo-posterior probability
\begin{equation}
    p(\mathcal{M}_i \mid D) = \frac{\exp\left(-\frac{1}{2} \Delta \mathrm{BIC}_i \right)}{\sum_j \exp\left(-\frac{1}{2} \Delta \mathrm{BIC}_j \right)},
\end{equation}
where $\Delta \mathrm{BIC}_i = \mathrm{BIC}_i - \mathrm{BIC}_\mathrm{min}$ is the relative BIC with respect to the best-scoring model. An example is shown in Fig.~\ref{fig:sm-closure}, illustrating the distribution of $p(\mathcal{M}_i \mid D)$ across the GA-selected models for a representative pseudodata replica. The SM clearly emerges as the most probable explanation, with other models receiving exponentially suppressed support. While one might be tempted to consider the collective weight of EFT-deformed models, such a sum lacks statistical meaning: each model represents a distinct hypothesis, and no prior has been defined over the full EFT space to justify their aggregation.

In summary, these results validate our framework by showing that it does not spuriously favor SMEFT deformations when applied to SM-consistent data, and maintains high specificity under realistic uncertainties. This provides a robust baseline for interpreting discovery-oriented tests with BSM injections.

\noindent {\bf Identifying imprints of New Physics.}
To validate the algorithm’s ability to identify genuine NP signals, we perform closure tests using pseudodata generated from a benchmark UV model. We consider as an example a heavy right-handed up-type quark $U$ with mass $M_U = 1~\mathrm{TeV}$ and coupling strength $\lambda_U = 1$, which induces a distinctive pattern of SMEFT contributions at the matching scale~\cite{terHoeve:2023pvs,deBlas:2017xtg}. Motivated by the SM flavor structure, we assume that $U$ couples to SM fermions via Yukawa-like interactions proportional to the SM Yukawas, therefore keeping only the third-generation coupling nonzero. As a result, the model generates operators involving exclusively third-generation quarks.  At dimension-6 and restricting to tree-level matching, this setup leads to three active SMEFT operators: 
\begin{equation}
    c_{\varphi Q}^{(3)} = -\frac{\lambda_U^2}{4 M_U^2}, \quad 
    c_{\varphi Q}^{(-)} = \frac{\lambda_U^2}{2 M_U^2}, \quad 
    c_{t \varphi} = y_t \frac{\lambda_U^2}{2 M_U^2} \,.
\end{equation}

Matching is performed at the scale $\mu_0 = M$, where the Wilson coefficients are initialized. The RGE is then used to evolve the coefficients to the relevant energy scales for each observable.
We note in passing that the choice of $M_U = 1~\mathrm{TeV}$ is made solely for illustrative purposes, as such a low mass is already excluded by direct searches. 

We apply the GA to pseudodata generated from the UV benchmark model under both ideal (L0) and fluctuated (L1) conditions. In the L0 scenario, the algorithm identifies the injected operators at both LHC and HL-LHC. However, only at the HL-LHC do the criteria consistently recover the full combination as the best model. In all L0 tests, the SM is clearly disfavored, confirming the algorithm’s sensitivity to genuine deformations.

The L1 scenario, which includes realistic statistical fluctuations, presents a more demanding test. Since AIC tends to overfit by favoring overly complex models even in the absence of signals, we exclude it from further analysis. Using BIC, the correct operators are generally among the most favored at both LHC and HL-LHC precision, though not always top-ranked. HL-LHC projections improve discrimination by reducing statistical noise. Nonetheless, operators like $\mathcal{O}_{t\varphi}$ remain difficult to isolate due to degeneracies with $\mathcal{O}_{\varphi G}$ and $\mathcal{O}_{tG}$ in Higgs production, while $\mathcal{O}_{\varphi t}$ can mimic left-handed currents~\cite{Celada:2024mcf}. These ambiguities are UV-dependent, and closure tests help identify regions of EFT space where such limitations are intrinsic.

To illustrate the algorithm’s behavior in the presence of a BSM signal, the left panel of Fig.~\ref{fig:bsm-closure} shows the pseudo-posterior probabilities $p(\mathcal{M}_i \mid D)$ for the highest-ranked models in a representative pseudodata replica.
At current LHC precision, the SM is correctly disfavored, but no clear statistical preference emerges among the  top-ranked models, which all have $\Delta\mathrm{BIC} \leq 1$ with respect to the best one, indicating limited ability to characterize the underlying NP signal at this level of precision.

On the other hand, at the HL-LHC, the most probable model recovers $\mathcal{O}_{\varphi Q}^{(3)}$ and $\mathcal{O}_{\varphi Q}^{(-)}$, but replaces the Yukawa-like operator $\mathcal{O}_{t\varphi}$ with the chromomagnetic $\mathcal{O}_{tG}$. The second-best model favors $\mathcal{O}_{\varphi G}$, while the true injected combination ranks third. This reflects the degeneracy among $\mathcal{O}_{t\varphi}$, $\mathcal{O}_{tG}$, and $\mathcal{O}_{\varphi G}$ in Higgs production, which limits full model recovery. Importantly, $\mathcal{O}_{\varphi Q}^{(3)}$ and $\mathcal{O}_{\varphi Q}^{(-)}$ appear in all top eight models, showing that these current-type operators are reliably identified despite fluctuations and competing directions. 
This underscores the importance of the HL-LHC’s enhanced precision in resolving ambiguities and identifying the underlying NP.

The right panel of Fig.~\ref{fig:bsm-closure} displays the 95\% C.L. intervals obtained from the same pseudodata replica for the three injected Wilson coefficients, along with $c_{tG}$ and $c_{\varphi G}$, comparing the outcome of a traditional global fit with that of our model-selection-based procedure in the HL-LHC scenario. In the global fit, all coefficients remain consistent with zero, failing to capture the injected BSM signal. In contrast, our top-ranked models consistently identify a non-zero subset aligned with the true deformation. This difference highlights a key limitation of global fits: when a signal aligns with a narrow direction in parameter space, its effect can be diluted by marginalization over many poorly constrained directions, making it harder to detect. Our approach addresses this by prioritizing minimal deformations with strong statistical support, enhancing sensitivity to viable NP scenarios that might otherwise go unnoticed.  It is worth emphasizing that global marginalized fits and the present approach serve complementary purposes. The former determine the allowed parameter ranges under the most general SMEFT hypothesis, while the latter aims to identify operator combinations that most effectively capture potential deviations from SM predictions.

Recovery of the correct operator combination varies notably across pseudodata replicas due to statistical fluctuations. However, when analyzing the L0 data at the HL-LHC, which can be interpreted as an average over many L1 replicas, we find that all three injected operators have marginal posterior probabilities\footnote{Marginal posterior probabilities are obtained via marginalization, which, in this model selection context, corresponds to summing the probabilities of all models that include the operator of interest.} of around $45-60\%$ each. Additionally, $\mathcal{O}_{\varphi t}$ ($\sim 50\%$), $\mathcal{O}_{\varphi G}$ ($\sim 17\%$), $\mathcal{O}_{tG}$ ($\sim 11\%$), $\mathcal{O}^{1}_{Qt}$ ($\sim 8\%$), and $\mathcal{O}_{\varphi D}$ ($\sim 6\%$) also receive notable support, reflecting residual correlations in the parameter space.

These findings highlight the importance of designing observables that are optimally sensitive to orthogonal EFT directions~\cite{GomezAmbrosio:2022mpm,Chen:2020mev,Chen:2023ind}, and suggest that future high-precision facilities such as the FCC-ee could play a decisive role in lifting such degeneracies and improving operator-level resolution~\cite{Celada:2024mcf,Allwicher:2024sso,Maura:2024zxz, Greljo:2025ggc}.

\noindent {\bf Application to real data.}
\label{sec:real-data}
\begin{figure}
    \centering
    \includegraphics[width=1.0\linewidth]{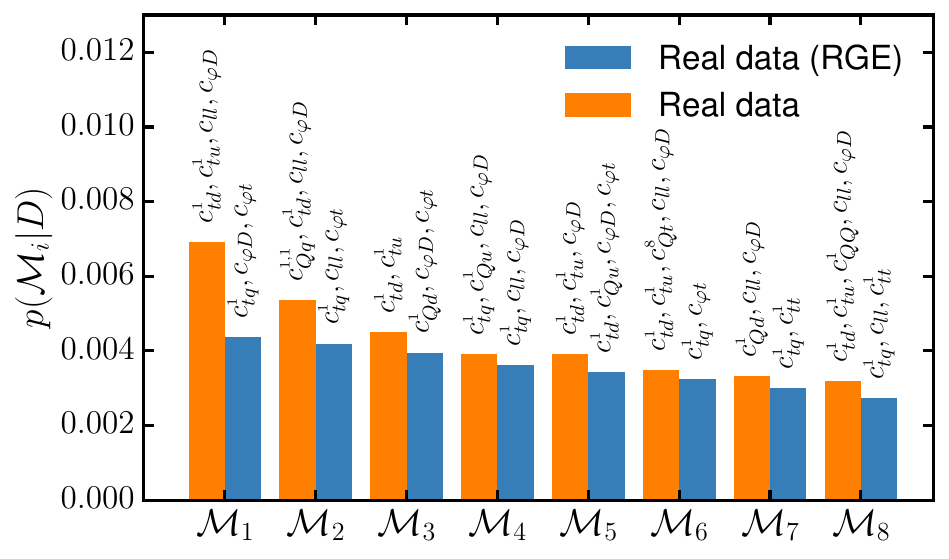}
    \caption{Same as Fig.~\ref{fig:sm-closure}, using real LEP + LHC data. Results are shown both with and without RGE, highlighting its impact on model selection.  Including RGE reduces the statistical preference for the top models, since operator mixing allows deviations to be accommodated by a broader set of SMEFT contributions. The SM does not appear in the top 8 models and is therefore not shown, but it remains statistically competitive and is not excluded.}
    \label{fig:real-data}
\end{figure}
We conclude by applying our methodology to the combined dataset of legacy LEP measurements and Higgs, top, and diboson data from LHC Run~2. Using the BIC as our model selection criterion, we find no compelling statistical evidence against the SM, which consistently ranks among the top-performing models and lies within $\Delta \mathrm{BIC} \approx 3$ of the best-scoring alternative. No specific SMEFT deformation yields a significantly improved fit once the complexity penalty is taken into account.

Nonetheless, models that outperform the SM (see Fig.~\ref{fig:real-data}) commonly feature color-singlet four-fermion operators involving top quarks, especially those relevant for $t\bar{t}$ production. Such operators are generally compatible with low-energy effects of heavy $Z'$ bosons or other neutral vector resonances preferentially coupled to third-generation quarks.  

While the current statistical evidence is limited, the repeated appearance of certain four-fermion operators in near-optimal models may suggest interesting directions for future study, though it could also stem from statistical fluctuations or residual mismodeling in the dataset. It is worth emphasizing that individual dimension-six SMEFT operators can originate from a wide variety of UV completions at tree level. To single out a particular UV resonance as the next step in the search for BSM physics will therefore require evidence from several operators. Mild deviations in $t\bar{t}$ kinematics have been investigated in the context of simplified model interpretations~\cite{ATLAS:2023gsl}, although any direct connection remains speculative at this stage.

\noindent {\bf Conclusions and Outlook.}
\label{sec:conclusions}
In this work, we introduced a new strategy for exploring the SMEFT parameter space, based on genetic algorithms and model selection. Our method identifies promising operator subsets that could explain deviations in data, without relying on UV assumptions. Compared to traditional global SMEFT analyses, this approach improves sensitivity to regions of parameter space where NP may manifest. While our focus has been on SMEFT, the methodology applies broadly to other EFTs and high-dimensional theory spaces where discovery is hindered by combinatorial complexity. Similar strategies could be used in low-energy EFTs for flavor transitions, neutrino non-standard interactions, or chiral perturbation theory, helping isolate minimal deformations consistent with data in the absence of strong UV guidance.

While our analysis has focused on synthetic data, applying this methodology to real experimental datasets will require careful treatment of several caveats. Present SMEFT fits often face inconsistencies across datasets and residual tensions unrelated to BSM effects. These challenges underscore the importance of cautious interpretation and the continued use of synthetic benchmarks to assess discovery thresholds and control false positives.

Furthermore, given the challenges encountered in disentangling correlated operators, a natural next step is to incorporate the squared contributions of SMEFT operators into the analysis, as these terms can help break degeneracies and improve parameter determination.

Looking ahead, the methodology developed here is well positioned to assess the discovery potential of future experimental programs at next-generation colliders. Facilities like FCC-ee, with their anticipated sub-permille precision on electroweak observables, offer an unprecedented opportunity to uncover subtle SMEFT effects. As the experimental frontier advances and datasets become increasingly precise and complex, algorithmic strategies such as the one here presented will be essential for navigating high-dimensional theory spaces and isolating faint signals of new physics.

\section*{Acknowledgements}
LM thanks Y. S. Koay for useful discussions on the use of genetic algorithms. LM thanks F. Maltoni and E. Vryonidou for their valuable comments on the manuscript.
LM acknowledges support from the European Union under the MSCA fellowship (Grant agreement N. 101149078) {\it Advancing global SMEFT fits in the LHC precision era (EFT4ward)}. This work is supported by the Spanish grants PID2023-147306NB-I00, PID2023-148162NB-C21, 
CEX2023-001292-S (MCIU/AEI/10.13039/501100011033), as well as CIPROM/2021/054 (Generalitat Valenciana).

\bibliography{genetic_eft}

\clearpage
\onecolumngrid

\appendix
\FloatBarrier
\begin{center}
\LARGE
{\bf End Matter}
\end{center}

\section{SMEFT operator definition}
\label{app:op-def}

In this appendix, we summarize the set of operators used in our analysis and provide their explicit definitions. The operator basis corresponds to the Warsaw basis of dimension-six operators~\cite{Grzadkowski:2010es}, subject to a $\text{U}(2)_q \times \text{U}(3)_d \times \text{U}(2)_u \times (\text{U}(1)_\ell \times \text{U}(1)_e)^3$ flavour symmetry assumption, and matches the implementation adopted in the \texttt{smefit} framework~\cite{Giani_2023}.

The purely bosonic operators are listed in Table~\ref{tab:oper_bos}, while the two-fermion and four-lepton operators are presented in Tables~\ref{tab:oper_ferm_bos} and~\ref{tab:oper_ferm_bos2}, respectively. The four-fermion operators are defined in Table~\ref{tab:oper_fourtop}.

%%%%%%%%%%%%%%%%%%%%%%%%%%%%%%%%%%%%%%%%%%%%%%%%%%%%%%%5
\begin{table}[H] 
  \begin{center}
    \renewcommand{\arraystretch}{1.6}
        \begin{tabular}{lll|lll}
          \toprule
          Operator $\quad$ & Coefficient $\quad$ & Definition& Operator $\quad$ & Coefficient $\quad$ & Definition \\
        \midrule
        \midrule
        $\Op{\varphi G}$ & $c_{\varphi G}$  & $\left( \varphi^\dagger \varphi - \frac{v^2}{2} \right)G^{\mu\nu}_{\sss A}\,
        G_{\mu\nu}^{\sss A}$ 
        & 
        $\Op{\varphi \square}$ & $c_{\varphi \square}$ & $(\pdp)\square(\pdp)$ \\
        %%%%%%%%%%%%%%%%%%%%%%%%%%%%%%%%%%%%%%%%%%%%%%%%%%%%%%%%%%%%%%%
        $\Op{\varphi B}$ & $c_{\varphi B}$ & $\left( \varphi^\dagger \varphi - \frac{v^2}{2} \right) B^{\mu\nu}\,B_{\mu\nu}$
        &
        $\Op{\varphi D}$ & $c_{\varphi D}$ & $(\varphi^\dagger D^\mu\varphi)^\dagger(\varphi^\dagger D_\mu\varphi)$ \\ 
        %%%%%%%%%%%%%%%%%%%%%%%%%%%%%%%%%%%%%%%%%%%%%%%%%%%%%%%%%%%%%%%
        $\Op{\varphi W}$ &$c_{\varphi W}$ & $\left( \varphi^\dagger \varphi - \frac{v^2}{2} \right)W^{\mu\nu}_{\sss I}\,
        W_{\mu\nu}^{\sss I}$ 
        &
        $\mathcal{O}_{W}$&   $c_{WWW}$ & $\epsilon_{IJK}W_{\mu\nu}^I W^{J,\nu\rho} W^{K,\mu}_\rho$ \\ 
        %%%%%%%%%%%%%%%%%%%%%%%%%%%%%%%%%%%%%%%%%%%%%%%%%%%%%%%%%%%%%%%
        $\Op{\varphi W B}$ &$c_{\varphi W B}$ & $(\varphi^\dagger \tau_{\sss I}\varphi)\,B^{\mu\nu}W_{\mu\nu}^{\sss I}\,$
        &
        $\Op{\varphi}$ & $c_{\varphi}$ & $\left( \varphi^\dagger \varphi - \frac{v^2}{2} \right)^3$ \\
        %%%%%%%%%%%%%%%%%%%%%%%%%%%%%%%%%%%%%%%%%%%%%%%%%%%%%%%%%%%%%%%%%%
       \bottomrule
        \end{tabular}
        \caption{Bosonic operators at dimension six that modify Higgs dynamics and electroweak gauge interactions.
          \label{tab:oper_bos}}
\end{center}
\end{table}
%%%%%%%%%%%%%%%%%%%%%%%%%%%%%%%%%%%%%%%%%%%%%%%%%%%%%%%%%%

\begin{table}[H]
  \begin{center}
    \renewcommand{\arraystretch}{1.45}
    \begin{tabular}{p{1.5cm} p{1.4cm} p{4.4cm} | p{1.5cm} p{1.4cm} p{4.5cm}}
      \toprule
      Operator & Coefficient & $\qquad$ Definition & Operator  & Coefficient & $\qquad$ Definition \\
                \midrule \midrule
      \multicolumn{6}{c}{3rd generation quarks} \\
                \midrule \midrule
%%%%%%%%%%%%%%%%%%%%%%%%%%%%%%%%%%%%%%%%%%%%%%%%%%%%%%%%%%%%%%%%%%%%%%%%%%%
    $\Op{\varphi Q}^{(1)}$ & $c_{\varphi Q}^{(1)}$~(*) & $i\big(\varphi^\dagger\lra{D}_\mu\,\varphi\big)
 \big(\bar{Q}\,\gamma^\mu\,Q\big)$ 
 &
 $\Op{tW}$ & $c_{tW}$ & $i\big(\bar{Q}\tau^{\mu\nu}\,\tau_{\sss I}\,t\big)\,
 \tilde{\varphi}\,W^I_{\mu\nu}
 + \text{h.c.}$ \\ 
%%%%%%%%%%%%%%%%%%%%%%%%%%%%%%%%%%%%%%%%%%%%%%%%%%%%%%%%%%%%%%%%%%%%%%%%%%%
    $\Op{\varphi Q}^{(3)}$ & $c_{\varphi Q}^{(3)}$  & $i\big(\varphi^\dagger\lra{D}_\mu\,\tau_{\sss I}\varphi\big)
 \big(\bar{Q}\,\gamma^\mu\,\tau^{\sss I}Q\big)$ 
 &
 $\Op{tB}$ & $c_{tB}$~(*) &
 $i\big(\bar{Q}\tau^{\mu\nu}\,t\big)
 \,\tilde{\varphi}\,B_{\mu\nu}
 + \text{h.c.}$\\ 
 %%%%%%%%%%%%%%%%%%%%%%%%%%%%%%%%%%%%%%%%%%%%%%%%%%%%%%%%%%%%%%%%%%%%%%%%%%%
    $\Op{\varphi t}$ & $c_{\varphi t}$& $i\big(\varphi^\dagger\,\lra{D}_\mu\,\,\varphi\big)
 \big(\bar{t}\,\gamma^\mu\,t\big)$
 &
  $\Op{t G}$ & $c_{tG}$ & $ig{\sss S}\,\big(\bar{Q}\tau^{\mu\nu}\,T_{\sss A}\,t\big)\,
 \tilde{\varphi}\,G^A_{\mu\nu}
 + \text{h.c.}$ \\ 
%%%%%%%%%%%%%%%%%%%%%%%%%%%%%%%%%%%%%%%%%%%%%%%%%%%%%%%%%%%%%%%%%%%%%%%%%%%
     $\Op{t \varphi}$ & $c_{t\varphi}$ & $\left(\pdp\right)
 \bar{Q}\,t\,\tilde{\varphi} + \text{h.c.}$ 
 &
  $\Op{b \varphi}$ & $c_{b\varphi}$ & $\left(\pdp\right)
 \bar{Q}\,b\,\varphi + \text{h.c.}$ \\  
%%%%%%%%%%%%%%%%%%%%%%%%%%%%%%%%%%%%%%%%%%%%%%%%%%%%%%%%%%%%%%%%%%%%%%%%%%%
                \midrule \midrule
                \multicolumn{6}{c}{1st, 2nd generation quarks} \\
                \midrule \midrule
%%%%%%%%%%%%%%%%%%%%%%%%%%%%%%%%%%%%%%%%%%%%%%%%%%%%%%%%%%%%%%%%%%%%%%%%%%%
    $\Op{\varphi q}^{(1)}$ & $c_{\varphi q}^{(1)}$~(*) & $\sum\limits_{\sss i=1,2} i\big(\varphi^\dagger\lra{D}_\mu\,\varphi\big)
 \big(\bar{q}_i\,\gamma^\mu\,q_i\big)$ 
 &
 ${\Op{\varphi d }}$ &
      ${{c_{\varphi d}}}$ & $\sum\limits_{\sss i=1,2,3} i\big(\varphi^\dagger\,\lra{D}_\mu\,\,\varphi\big)
 \big(\bar{d}_i\,\gamma^\mu\,d_i\big)$\\ 
%%%%%%%%%%%%%%%%%%%%%%%%%%%%%%%%%%%%%%%%%%%%%%%%%%%%%%%%%%%%%%%%%%%%%%%%%%%
    $\Op{\varphi q}^{(3)}$ & $c_{\varphi q}^{(3)}$ & $\sum\limits_{\sss i=1,2} i\big(\varphi^\dagger\lra{D}_\mu\,\tau_{\sss I}\varphi\big)
 \big(\bar{q}_i\,\gamma^\mu\,\tau^{\sss I}q_i\big)$
 &$\Op{c \varphi}$ & $c_{c \varphi}$ & $\left(\pdp\right)
 \bar{q}_2\,c\,\tilde\varphi + \text{h.c.}$ \\ 
%%%%%%%%%%%%%%%%%%%%%%%%%%%%%%%%%%%%%%%%%%%%%%%%%%%%%%%%%%%%%%%%%%%%%%%%%%%
  ${\Op{\varphi u }}$ &
      ${{c_{\varphi u}}}$ & $\sum\limits_{\sss i=1,2} i\big(\varphi^\dagger\,\lra{D}_\mu\,\,\varphi\big)
 \big(\bar{u}_i\,\gamma^\mu\,u_i\big)$\\ 
    
 %%%%%%%%%%%%%%%%%%%%%%%%%%%%%%%%%%%%%%%%%%%%%%%%%%%%%%%%%%%%%%%%%%%%%%%%%%%
                \midrule \midrule
		      \multicolumn{6}{c}{two-leptons} \\
                \midrule \midrule
%%%%%%%%%%%%%%%%%%%%%%%%%%%%%%%%%%%%%%%%%%%%%%%%%%%%%%%%%%%%%%%%%%%%%%%%%%%
    $\Op{\varphi \ell_i}$ & $c_{\varphi \ell_i}$ & $ i\big(\varphi^\dagger\lra{D}_\mu\,\varphi\big)
   \big(\bar{\ell}_i\,\gamma^\mu\,\ell_i\big)$ 
   &
    $\Op{\varphi \mu}$ & $c_{\varphi \mu}$ & $ i\big(\varphi^\dagger\lra{D}_\mu\,\varphi\big)
 \big(\bar{\mu}\,\gamma^\mu\,\mu\big)$ \\  
%%%%%%%%%%%%%%%%%%%%%%%%%%%%%%%%%%%%%%%%%%%%%%%%%%%%%%%%%%%%%%%%%%%%%%%%%%%
    $\Op{\varphi \ell_i}^{(3)}$ & $c_{\varphi \ell_i}^{(3)}$ & $ i\big(\varphi^\dagger\lra{D}_\mu\,\tau_{\sss I}\varphi\big)
 \big(\bar{\ell}_i\,\gamma^\mu\,\tau^{\sss I}\ell_i\big)$ 
 &
  $\Op{\varphi \tau}$ & $c_{\varphi \tau}$ & $i\big(\varphi^\dagger\lra{D}_\mu\,\varphi\big)
 \big(\bar{\tau}\,\gamma^\mu\,\tau\big)$ \\  
%%%%%%%%%%%%%%%%%%%%%%%%%%%%%%%%%%%%%%%%%%%%%%%%%%%%%%%%%%%%%%%%%%%%%%%%%%%
    $\Op{\varphi e}$ & $c_{\varphi e}$ & $ i\big(\varphi^\dagger\lra{D}_\mu\,\varphi\big)
 \big(\bar{e}\,\gamma^\mu\,e\big)$ 
 &
  $\Op{\tau \varphi}$ & $c_{\tau \varphi}$ & $\left(\pdp\right)
 \bar{\ell_3}\,\tau\,{\varphi} + \text{h.c.}$ \\
%%%%%%%%%%%%%%%%%%%%%%%%%%%%%%%%%%%%%%%%%%%%%%%%%%%%%%%%%%%%%%%%%%%%%%%%%%%
                \midrule \midrule
		      \multicolumn{6}{c}{four-leptons} \\
                \midrule \midrule
 $\Op{\ell\ell}$ & $c_{\ell\ell}$ & $\left(\bar \ell_1\gamma_\mu \ell_2\right) \left(\bar \ell_2\gamma^\mu \ell_1\right)$ 
 &
 $\Op{\ell\ell}^{11}$ & $c^{11}_{\ell\ell}$ & $\left(\bar \ell_1\gamma_\mu \ell_1\right) \left(\bar \ell_1 \gamma^\mu \ell_1\right)$
 \\
 
  \bottomrule
\end{tabular}
\caption{Same as Table~\ref{tab:oper_bos}, but for operators containing two fermion fields, either quarks or leptons, as well as the four-lepton operator $\mathcal{O}_{\ell\ell}$ and $\mathcal{O}^{11}_{\ell\ell}$. The flavour index $i$ runs from 1 to 3. Coefficients marked with an asterisk (*) in the second column do not correspond to independent physical degrees of freedom in the fit, but are instead replaced by $c_{\varphi q}^{(-)}$, $c_{\varphi Q}^{(-)}$, and $c_{tZ}$, as defined in Table~\ref{tab:oper_ferm_bos2}.
\label{tab:oper_ferm_bos}}
\end{center}
\end{table}
%%%%%%%%%%%%%%%%%%%%%%%%%%%%%%%%%%%%%%%%%%%%%%%%%%%%%%%

\begin{table}[H]
  \begin{center}
    \renewcommand{\arraystretch}{1.45}
    \begin{tabular}{l l}
    \toprule
     Coefficient $\qquad$ & Definition\\ \hline
     \midrule
    %%%%%%%%%%%%%%%%%%%%%%%%%%%%%%
    $c_{\varphi Q}^{(-)}$ & $c_{\varphi Q}^{(1)}-c_{\varphi Q}^{(3)}$\\ 
    %%%%%%%%%%%%%%%%%%%%%%%%%%%%%%%%%
    $c_{tZ}$ & $-s_\theta \, c_{tB}+ c_\theta \,c_{tW}$ \\ 
    %%%%%%%%%%%%%%%%%%%%%%%%%%%%%%%%%%
    $c_{\varphi q}^{(-)}$ & $c_{\varphi q}^{(1)}-c_{\varphi q}^{(3)}$ \\ 
    \bottomrule
\end{tabular}
\caption{Linear combinations of the two-fermion operators listed in Table~\ref{tab:oper_ferm_bos}, which replace those marked with an asterisk (*) at the fit level.
\label{tab:oper_ferm_bos2}}
\end{center}
\end{table}

%%%%%%%%%%%%%%%%%%%%%%%%%%%%%%%%%%%%%%%%%%%%%%%%%%%%%%%5
\begin{table}[H] 
  \begin{center}
    \renewcommand{\arraystretch}{1.53}
        \begin{tabular}{ll| ll}
          \toprule
          DoF $\qquad$ &  Definition (in  Warsaw basis notation) & DoF $\qquad$ &  Definition (in  Warsaw basis notation) \\
          \midrule
          \midrule
      $c_{QQ}^1$    &   $2\ccc{1}{qq}{3333}-\frac{2}{3}\ccc{3}{qq}{3333}$ 
      &
      $c_{QQ}^8$       &         $8\ccc{3}{qq}{3333}$\\  
%%%%%%%%%%%%%%%%%%%%%%%%%%%%%%%%%%5%%%%%%%%%%%%%%%%%%%%%%%%%%%%%%%%%%%%%%%%%%%%%%%%%%%%%%%%%%%%%%%%%%%%%%
     $c_{Qt}^1$         &         $\ccc{1}{qu}{3333}$
     &
     $c_{Qt}^8$         &         $\ccc{8}{qu}{3333}$\\   
%%%%%%%%%%%%%%%%%%%%%%%%%%%%%%%%%%5%%%%%%%%%%%%%%%%%%%%%%%%%%%%%%%%%%%%%%%%%%%%%%%%%%%%%%%%%%%%%%%%%%%%%%
            \midrule      
%%%%%%%%%%%%%%%%%%%%%%%%%%%%%%%%%%5%%%%%%%%%%%%%%%%%%%%%%%%%%%%%%%%%%%%%%%%%%%%%%%%%%%%%%%%%%%%%%%%%%%%%%
  $c_{Qq}^{1,8}$       &  	 $\ccc{1}{qq}{i33i}+3\ccc{3}{qq}{i33i}$  
  &
  $c_{Qq}^{1,1}$         &   $\ccc{1}{qq}{ii33}+\frac{1}{6}\ccc{1}{qq}{i33i}+\frac{1}{2}\ccc{3}{qq}{i33i} $   \\    
 %%%%%%%%%%%%%%%%%%%%%%%%%%%%%%%%%%5%%%%%%%%%%%%%%%%%%%%%%%%%%%%%%%%%%%%%%%%%%%%%%%%%%%%%%%%%%%%%%%%%%%%%%
   $c_{Qq}^{3,8}$         &   $\ccc{1}{qq}{i33i}-\ccc{3}{qq}{i33i} $  
   &
$c_{Qq}^{3,1}$          & 	$\ccc{3}{qq}{ii33}+\frac{1}{6}(\ccc{1}{qq}{i33i}-\ccc{3}{qq}{i33i}) $   \\     
 %%%%%%%%%%%%%%%%%%%%%%%%%%%%%%%%%%5%%%%%%%%%%%%%%%%%%%%%%%%%%%%%%%%%%%%%%%%%%%%%%%%%%%%%%%%%%%%%%%%%%%%%%
$c_{tq}^{8}$         &  $ \ccc{8}{qu}{ii33}   $ 
  &
$c_{tq}^{1}$       &   $  \ccc{1}{qu}{ii33} $\\   
 %%%%%%%%%%%%%%%%%%%%%%%%%%%%%%%%%%5%%%%%%%%%%%%%%%%%%%%%%%%%%%%%%%%%%%%%%%%%%%%%%%%%%%%%%%%%%%%%%%%%%%%%%
$c_{tu}^{8}$      &   $2\ccc{}{uu}{i33i}$ 
 &
$c_{tu}^{1}$        &   $ \ccc{}{uu}{ii33} +\frac{1}{3} \ccc{}{uu}{i33i} $ \\   
 %%%%%%%%%%%%%%%%%%%%%%%%%%%%%%%%%%5%%%%%%%%%%%%%%%%%%%%%%%%%%%%%%%%%%%%%%%%%%%%%%%%%%%%%%%%%%%%%%%%%%%%%%
$c_{Qu}^{8}$         &  $  \ccc{8}{qu}{33ii}$
 &
 $c_{Qu}^{1}$     &  $  \ccc{1}{qu}{33ii}$  \\    
 %%%%%%%%%%%%%%%%%%%%%%%%%%%%%%%%%%5%%%%%%%%%%%%%%%%%%%%%%%%%%%%%%%%%%%%%%%%%%%%%%%%%%%%%%%%%%%%%%%%%%%%%%
 $c_{td}^{8}$        &   $\ccc{8}{ud}{33jj}$ 
  &
 $c_{td}^{1}$          &  $ \ccc{1}{ud}{33jj}$ \\    
  %%%%%%%%%%%%%%%%%%%%%%%%%%%%%%%%%%5%%%%%%%%%%%%%%%%%%%%%%%%%%%%%%%%%%%%%%%%%%%%%%%%%%%%%%%%%%%%%%%%%%%%%%
 $c_{Qd}^{8}$        &   $ \ccc{8}{qd}{33jj}$ 
 &
 $c_{Qd}^{1}$         &   $ \ccc{1}{qd}{33jj}$\\
        %%%%%%%%%%%%%%%%%%%%%%%%%%%%%%%%%%5%%%%%%%%%%%%%%%%%%%%%%%%%%%%%%%%%%%%%%%%%%%%%%%%%%%%%%%%%%%%%%%%%%%%%%
         \bottomrule
  \end{tabular}
  \caption{Definitions of the four-fermion coefficients entering the fit. The coefficients are grouped into two categories: four-heavy operators (upper part) and two-light–two-heavy operators (lower part). The flavour indices $i$ and $j$ take values $i = 1, 2$ and $j = 1, 2, 3$, respectively.
\label{tab:oper_fourtop}}
  \end{center}
\end{table}
%%%%%%%%%%%%%%%%%%%%%%%%%%%%%%%%%%%%%%%%%%%%%%%%%%%%%%%%%%%%%%%%%%%%%%%%%%%%%%%

\section{Genetic Algorithm Implementation}
\label{app:ga}

In this appendix, we delineate the specifics of the implementation of the genetic algorithm (GA) employed in this work. 
Each candidate SMEFT model explored in our analysis is represented as a binary string of length 
$N_{\text{op}} = 52$, corresponding to the set of CP-even dimension-six operators included in the fit. 
A bit value of~1~(0) indicates that the associated operator is active~(inactive) in the model, 
so the genetic algorithm samples combinations of operator \emph{presence} only. 
No functional relations among Wilson coefficients are assumed: whenever an operator is included, 
its corresponding Wilson coefficient is treated as a free continuous parameter optimized through 
$\chi^2$ minimization within the \texttt{smefit} framework. 
The genetic algorithm thus performs a combinatorial search over model structures, 
not over the numerical values of the coefficients themselves.

The population is evolved through standard GA operations: 
(i)~\textbf{selection}, implemented via tournament selection with $K = 4$; 
(ii)~\textbf{crossover}, where two parent strings exchange bits under a uniform scheme; 
and (iii)~\textbf{mutation}, where each bit flips with probability $p_{\text{mut}} = 0.025$. 
The initial population of 2000 candidate models includes by construction the Standard Model and all single-operator hypotheses, a choice found to enhance both stability and exploration efficiency. This population is then evolved for 200 generations, with the 100 fittest individuals selected as parents at each step.
Fitness is defined through the Bayesian Information Criterion (BIC) or Akaike Information Criterion (AIC), 
which balance quality of fit and model complexity. 
The algorithm therefore searches the discrete model space $\{0,1\}^{52}$ 
for subsets of operators that maximize statistical support.

It is worth emphasizing that, unlike genetic programming or symbolic regression, 
where the nodes of the evolutionary tree represent mathematical operations 
(e.g.\ addition, multiplication, or trigonometric functions), 
our genetic algorithm does not evolve analytical expressions. 
The SMEFT Lagrangian already fixes the functional form of each operator’s contribution, 
and the GA merely determines which operators are switched on or off in a given model. 
The “nodes” in our case correspond to binary inclusion variables, 
not to algebraic operations, making the search purely combinatorial 
over operator subsets rather than functional expressions.
\end{document}